\documentclass[aps,prl,preprint,groupedaddress]{revtex4}

\usepackage{graphicx}
\usepackage{amsmath,amssymb,dsfont}
\usepackage{dcolumn}

\newcommand{\beq}{\begin{eqnarray}}
\newcommand{\eeq}{\end{eqnarray}}

\begin{document}
\title{Bosonic hard spheres  
in quasi-one dimensional bichromatic optical lattices}
\author{M.C. Gordillo, C. Carbonell-Coronado and F.~De~Soto}
\affiliation{Departamento de Sistemas F\'{\i}sicos Qu\'{\i}micos y Naturales,
Universidad Pablo de Olavide. 41013 Sevilla, Spain}

\begin{abstract}
We calculated the phase diagram of a continuous system of hard spheres 
loaded in a quasi-one dimensional bichromatic optical lattice. 
The wavelengths of both lattice-defining lasers were 
chosen to model an incommensurate arrangement. Densities of one particle and half a particle per 
potential well were considered. Our results 
can be compared directly to those of the experimental system [Fallani et al. 
PRL, {\bf 98} 130404 (2007)] from which our initial parameters were taken.  
The phase diagrams for both densities are significatively different to those obtained by describing 
the same experimental setup with a Bose-Hubbard model.  

\end{abstract}


\maketitle
\section{Introduction}
An optical lattice is the result of the interference of one or several pairs of laser beams to produce a stationary wave,
felt by a neutral atom as an external potential of the form \cite{bloch1,bloch2}: 
\begin{eqnarray} \label{pot}
V_{ext}(x,y,z) = & & V_x \sin^2 \left( \frac{2 \pi x} {\lambda_x} \right) +  
 V_y \sin^2 \left( \frac{2 \pi y}{\lambda_y} \right)   \nonumber \\
&+& V_z \sin^2 \left(\frac{2 \pi z}{\lambda_z} \right) .
\end{eqnarray}
This expression, that depends on the laser wavelengths ($\lambda_x,\lambda_y,\lambda_z$), defines a three dimensional lattice.      
$V_x,V_y$,$V_z$ are the depths of the potential barriers felt by the atoms and can be modified by changing the light intensity. In a very common arrangement, 
two of those depths are fixed to values high enough to produce a tight confinement in two directions. We have then quasi-one
dimensional systems well described by the continuous Hamiltonian:  
\begin{equation}\label{hamiltonian}
H = \sum_{i=1}^N \left[ -\frac{\hbar^2}{2m} \triangle  + V_{ext} (z_i) + \frac{1}{2} m \omega_\perp^2 (x_i^2 + y_i^2) \right]  + \sum_{i<j} V(r_{ij}).
\end{equation}
Here, $ V(r_{ij})$ represents the interaction between the atoms $i$ and $j$, separated a distance $r_{ij}$ from each other, and  
$m$ is the mass of those atoms. $\omega_\perp$ is the harmonic frequency we obtain at developing the first two terms of the right-hand side of 
Eq. (\ref{pot}) around $x=0$ and $y=0$, a good approximation when $V_x$ and $V_y$ are deep enough.
When the interaction between particles is modeled by a hard spheres one, a system described 
by this Hamiltonian  at a density of one particle per potential well exhibits superfluid-Mott insulator transitions \cite{gre2} for values of $V_z$ that depend 
on $\omega_\perp$ \cite{feli3}. Eq. (\ref{hamiltonian}) does not consider the system to be harmonically trapped in the longitudinal direction, $z$. That would 
mean to include an additional term of the form $\frac{1}{2} m \omega_z^2 z_i^2$ in the Hamiltonian \cite{feli5}.

Disorder can be introduced on this simple arrangement. Two are the most common experimental routes: the use of two superimposed lasers of different (and
incommensurate) wavelengths on 
the $z$ direction \cite{fallani,roati,gadway,prl113}, or the introduction of a speckle field on the optical lattice \cite{white,pasienski}. In both cases, 
we end up having potential wells of unequal depths, with a quasi-periodic variation in the first case and a more randomly distribution of potential minima
in the second.  
What the phase diagrams of those arrangements have in common is the presence of at least a superfluid (SF) and a Bose glass (BG) phase, 
depending on the particular experimental setup. If the average number of atoms per potential well is an integer, a Mott insulator (MI)
can be also stable for some values of the lattice defining parameters \cite{fisher}. 

In this work, we considered the first disorder option, and solved the Schr\"odinger equation corresponding to the continuous Hamiltonian in Eq. (\ref{hamiltonian}) with  
an optical lattice potential of the form:
\begin{equation}\label{fa}
V_{ext}(z) = s_1 E_{R_1} \sin^2 \left( \frac{2 \pi z} {\lambda_1} \right) + s_2 E_{R_2} \sin^2 \left( \frac{2 \pi z} {\lambda_2} \right) 
\end{equation}
where $\lambda_1$ and $\lambda_2$ are the two wavelengths that define the bichromatic lattice. Here, E$_{R_1}$ and  E$_{R_2}$ are the recoil energies 
corresponding to $\lambda_1$ and $\lambda_2$ (E$_{R_x}$= $h^2/(2m \lambda_x^2$)). 
Since we chose to reproduce as closely as possible the experimental    
setup used in Ref. \onlinecite{fallani}, we took $\lambda_1$ = 830 nm,  $s_{\perp}$ = 40 (in units of  E$_{R_1}$), and $m$ as the mass of a $^{87}$Rb atom.    
There is a difference, tough. We considered $\lambda_2$ = 1067 nm instead of the 1076 nm originally used. This makes the ratio $\lambda_1/\lambda_2$ = $7/9 \sim 0.7778$
instead of the 0.77137 of the experimental data. $s_1$ and $s_2$ where taken to be in the same range as the experimental data ($s_1 =$ 16; 0 $\le s_2 \leq$ 2.5, in units of their
respective recoil energies). 
To use rational approximations to  the real wavelength ratio is a common practice
(see for instance Ref. \onlinecite{roth}, in which $\lambda_1/\lambda_2$ = $5/7$, or Ref. \onlinecite{roscilde} in 
which $\lambda_1/\lambda_2$ = $97/126$), and allowed us to use relatively small simulation cells. 
The interatomic potential 
was described by a hard-spheres (HS) interaction:  $V(r_{ij})=+\infty$ for $r_{ij}<a$ and 0 otherwise,  
$a$ being fixed to a single value, that of the scattering length of the  
 $^{87}$Rb atoms \cite{pethick}. 
This potential was previously used   
to describe bosons loaded in optical lattices \cite{feli3,feli1,pilati,feli2,feli4,feli5}.

In most previous literature, the continuous Hamiltonian of Eq.(\ref{hamiltonian}) is approximated by the discrete Bose-Hubbard (BH) model \cite{jaksch}: 
\begin{equation}
H=-J \sum_{<ij>} b_i^+ b_j+ \frac{U}{2} \sum_i n_i(n_i-1) +  \sum_i \Delta_i n_i,
\label{hub}
\end{equation}
in which all the particles are allowed to stay only at the minima of their corresponding optical lattice potentials ($i$ sites).    
Besides, only interactions of the type  $<ij>$ were considered, $j$ being one of the two nearest neighbors (a quasi-one dimensional system translates into a 
pure one-dimensional Hamiltonian) to any particular $i$.   
$b^+_i(b_i)$ is the bosonic creation (annihilation) operator at 
$i$, and $n_i$ represents the number of neutral atoms at that same location.  
$J$, $U$ and $\epsilon_i$ are parameters that can be related meaningfully to the experimental ones ($V_z$, $\lambda$, and $\omega_{\perp}$) only  when the optical lattice 
potentials are deep enough \cite{bloch2,jaksch}. For the kind of Hamiltonians we are dealing with in this work \cite{bloch2}: 
\begin{equation}\label{J}
 J=\frac{4}{\sqrt\pi}E_{R}\left(\frac{V_z}{E_R}\right)^{3/4}e^{-2\sqrt{\frac{V_z}{E_R}}},
\end{equation}
and
\begin{equation}\label{U}
 U=\sqrt{\frac{2}{\pi}}\hbar\omega_{\perp}\left(\frac{V_z}{E_R} \right)^{1/4}\frac{2\pi}{(\lambda/a)}.
\end{equation}
$J$ is the hopping matrix element between sites $i$ and $j$, and $U$ models the repulsive interaction between two atoms located at the same $i$ site. To deduce
the expression for $U$, it was supposed that the full interaction potential between atoms  
in the original continuous Hamiltonian was adequately described by a pseudopotential \cite{bloch2}. Eqs. (\ref{J}) and (\ref{U}) were deduced for monochromatic lattices, 
i.e., lattices defined by a single laser of wavelength $\lambda$, and recoil energy $E_R$. If another laser is introduced, both expressions  would have to be recalculated
to include the effect of this second wavelength \cite{roux}. However, this is usually not done \cite{roth,roscilde,roux,deng,batrouni,kisker,gurarie}, even though 
a variation in $J$ could produce an appreciable change in the phase diagram. \cite{sengupta}   

The remaining parameter in Eq.(\ref{hub}), $\Delta_i$, represents the energy offset at each potential well, and for a monochromatic lattice with no longitudinal confinement 
is the same for each site. This means that could be set to zero without loss of generality. 
However, when disorder is introduced, this is no longer the case. For bichromatic lattices, $\Delta_i \propto \cos [2 \pi (\lambda_1/\lambda_2) i]$  \cite{prl113,roth,deng},
while for random disorder, $\Delta_i$ follows a uniform distribution in the interval $[-\Delta,0]$ or $[-\Delta,\Delta]$ \cite{batrouni,kisker,gurarie}.      

In this work, we calculate the phase diagram of the continuous Hamiltonian defined by Eq. (\ref{hamiltonian}) with the optical lattice potential of Eq. (\ref{fa}). This implies that 
the disorder arises naturally at the introduction of the second laser, with no {\em ad hoc} change in the depths of the potential wells. The only approximation  
was to use a hard spheres interparticle interaction, a very common approach \cite{feli3,feli1,pilati,feli2,feli5,feli5}. We considered two densities, one particle per optical 
lattice minimum, and half a particle per minimum, and compared our results to those calculated with a Bose-Hubbard model for the same systems \cite{roux}. However, our 
primary goal will be to compare our results to the experimental ones, when available. Since the lattice defined 
by $\lambda_1$ is in most cases deeper than the defined by $\lambda_2$ ($s_1 E_{R_1}$ $>$ $s_2 E_{R_2}$), to define the average density we used the number of potential minima corresponding to 
the lattice defined by $\lambda_1$. This convention will not affect the limits of the phase diagram, since when $s_1 E_{R_1} \sim s_2 E_{R_2}$, the system is always a superfluid, as we will see below.          
 
\section{Method}

The initial step to calculate the phase diagrams was to obtain the ground state of a set of atoms described by the Hamiltonian of Eq. (\ref{hamiltonian}).  
This means that we are operating a T=0, a reasonable approximation commonly used \cite{roth,roscilde,roux} to describe the very low temperature regimes in which the experiments are performed \cite{fallani,prl113}. 
In any case, we expect our description should be valid for temperatures lower than the corresponding to the excitations of the system, which are in the range of the kHz \cite{fallani}.  
To do so, we used the diffusion Monte Carlo (DMC) method \cite{boro94}. That technique needs an initial approximation to the ground state wavefunction to work, the  
so-called {\em trial} function. That initial expression depends on the positions of the $N$ atoms in the simulation cell ({\bf r$_1$},$\cdots$,{\bf r$_N$}). In this work, we used:  
\begin{equation} \label{trialtot}
\Phi({\bf r_1},\cdots,{\bf r_N}) = \prod_{i=1}^N \psi (x_i,y_i) \prod_{j=1}^N \phi (z_i) \prod_{l<m=1}^N \Psi(r_{lm})
\end{equation}
where $x_i,y_i,z_i$ are the coordinates of the atom $i$ and $r_{lm}$ is the distance between atoms $l$ and $m$.  
We took $\psi (x_i,y_i)$ to be a Gaussian of variance $\sigma_{\perp}^2 = \hbar/(m \omega_{\perp})$, i.e., the exact solution to the harmonic potential that traps the particles transversally.  
$\phi (z_i)$ was taken to be the numerical solution to an one-dimensional Schr\"odinger equation including only the monochromatic lattice defined by $\lambda_1$. The two body correlations were modeled by:   
\cite{boro1,feli4,feli5}:
\begin{equation}\label{trial}                                    
\Psi(r_{ij})=\left\{
\begin{array}{lr}
0 &  r_{ij} < a\\
B \frac{\sin(\sqrt{\epsilon} (r_{ij}-a))}{r_{ij}} & a<r_{ij}<D,  \\ 
1-A e^{-r_{ij}/\gamma} & r_{ij}>D.
\end{array} 
\right. 
\end{equation}
$a$ is here the scattering length of the pair of atoms. The particular way to obtain the five constants in this equation has been detailed elsewhere \cite{feli4,feli5}. 

As indicated above, we chose the $s_1$ and $s_2$ parameters in the same range that the experimental ones of Ref. \onlinecite{fallani}. The main difference is that instead of limiting
ourselves to the case $s_1/E_{R_1}$ = 16, we considered the entire range 0 $ \le s_1/E_{R_1} \le$ 16. For each $s_1/E_{R_1}$ we performed calculations in the interval  0 $\le s_2/E_{R_2} \le$ 2.5 for a density of one atom 
per potential well and  0 $\le s_2/E_{R_2} \le$ 2 when the average atom occupation was 0.5. In most cases, a simulation cell including 36 potential minima (= 18 $\lambda_1$) was used. We checked that 
to increase that number to 54, 72 or even 90 did not vary the results presented in this work.       

\section{Results}

\subsection{One particle per potential well} 

A Bose glass is usually defined as an insulating (non-superfluid) and finite-compressibility (gapless) phase \cite{fisher,batrouni}. On the other hand, 
a Mott insulator is a non-superfluid gaped state.  So, to determine whether a 
particular arrangement of atoms can be described as a glass, we should calculate the superfluid fraction (to know if we are dealing with an insulator) and the compressibility of the system  
(to see if we have a MI or a BG). 
We define the compressibility, $\kappa$, as:  
\begin{equation}
\kappa = \frac{\partial n}{\partial \mu},
\label{kappa}
\end{equation}
where $\mu$ is the chemical potential, 
\begin{equation}
\mu = \frac{\partial E}{\partial N}
= n \frac{\partial \frac{E}{N}}{\partial n} + \frac{E}{N}.
\label{mudef}
\end{equation}
Here, $E$ is the total energy of the hard spheres arrangement loaded in the quasi-one dimensional optical lattice and $n$ is the density of the system 
($n=N/L$, $N$, number
of atoms; $L$, length of the simulation cell). 
This means that a finite $\kappa$ at a given $n$ is associated with a continuous slope of the energy per atom versus the density at that $n$. 

\begin{figure}
\begin{center}
\includegraphics[width=0.44\textwidth]{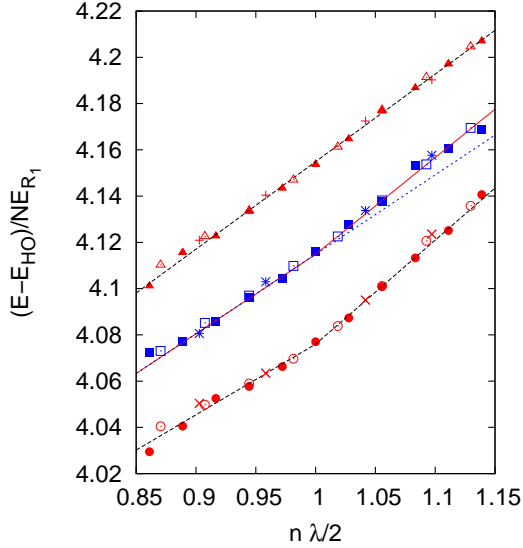} 
\caption{(Color online) 
Energy per particle (minus the harmonic contribution, $E_{HO}/N$) for $s_1$ = 16 in units of $E_{R_1}$ versus the filling fraction. 
Symbols correspond to the simulation results for different values of $s_2$ (from bottom to top s$_2$ = 1.2, s$_2$ = 1.3 and $s_2$ = 1.4, all of them in units of $E_{R_2}$). Full symbols correspond to 
simulation cells with 36 potential wells, open symbols to cells with 54 minima and crosses display the case for 72 wells.  
Lines correspond to least squares fits to the simulation data 
considering 
only the cell with 36 potential minima.
}
\label{fig1}
\end{center}
\end{figure}

In Fig. \ref{fig1}, we represent $E/N$, in units of the recoil energy of the primary lattice 
versus its filling fraction ($n \lambda_1$/2), subtracting from the energy the constant contribution per atom due to the transversal harmonic confinement, $E_{HO}/N$, in the same units.   
All the cases displayed correspond to $s_1/E_{R_1}$ = 16 and different values for $s_2$: $s_2/E_{R_2}$ = 1.2 (bottom), $s_2/E_{R_2}$ = 1.3 (middle), and  $s_2/E_{R_2}$ = 1.4 (top). 
To see the influence of the size of the simulation cell in our results, we display there the energies corresponding to simulation cells including 36 (full symbols), 
54 (open symbols) and 72 (crosses) potential minima. The results for bigger cells were similar and are not shown for simplicity.   
The dashed and full lines are least-squares fits to the simulation
results for densities below and above $n \lambda_1$/2 = 1 when only the data for 36 wells are considered. 
The fact that all the symbols are basically on top of those fits, indicates that our energies are not affected by size effects.  
The dotted line corresponds to the continuation of the result in the 0.85 $\le$ $n \lambda_1$/2 $\le$ 1 range beyond
that last density.
That line helps us to see that for $s_2/E_{R_2}$ = 1.3, the slope of the energy versus the filling fraction is discontinuous at $n \lambda_1$/2 = 1, something similar to what happens for
$s_2/E_{R_2}$ = 1.2. 

\begin{figure}
\begin{center}
\includegraphics[width=0.44\textwidth]{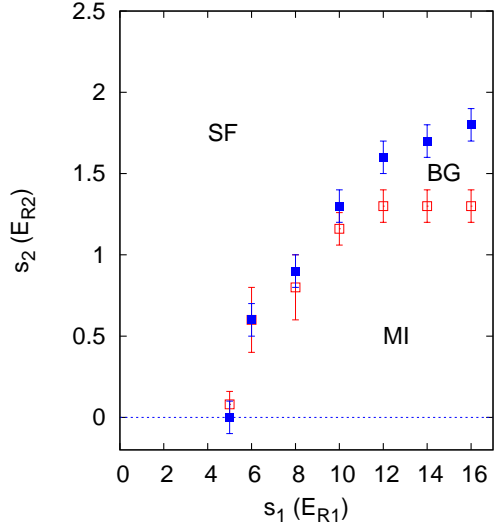} 
\caption{(Color online) 
Phase diagram for the case of an average density of one particle per potential well. SF, superfluid; BG, Bose glass; MI, Mott insulator.   
$s_1$ and $s_2$ are the parameters defined in Eq.(\ref{fa}), in units of their respective recoil energies. See further explanation in the text.
}
\label{fig2}
\end{center}
\end{figure}

On the other hand, when $s_2/E_{R_2}$ = 1.4 there is no such a slope change, what indicates that for those parameters we have a gapless phase. Using that information, and since
we considered the smallest difference between consecutive $s_2/E_{R_2}$ values to be 0.1, we estimated
the upper limit of the Mott insulator for $s_1/E_{R_1}$ = 16 to be  $s_2/E_{R_2}$ = 1.3 $\pm$ 0.1. This corresponds to the rightmost open symbol in Fig. \ref{fig2}.
For larger $s_2/E_{R_2}$ values, we have a gapless arrangement that can be either a superfluid or a Bose glass. That value is in reasonable agreement with the experimental results of
Ref. \onlinecite{fallani}, in which we can see a disappearance of the MI somewhere between $s_2/E_{R_2}$ = 0.5 and $s_2/E_{R_2}$ = 1.5. 
The remaining open symbols in Fig. \ref{fig2} were obtained in the same way 
that the already described for $s_1/E_{R_1}$ = 16 and indicate the upper limit of stability of the Mott insulator phase for each $s_1/E_{R_1}$ value.    

\begin{figure}
\begin{center}
\includegraphics[width=0.44\textwidth]{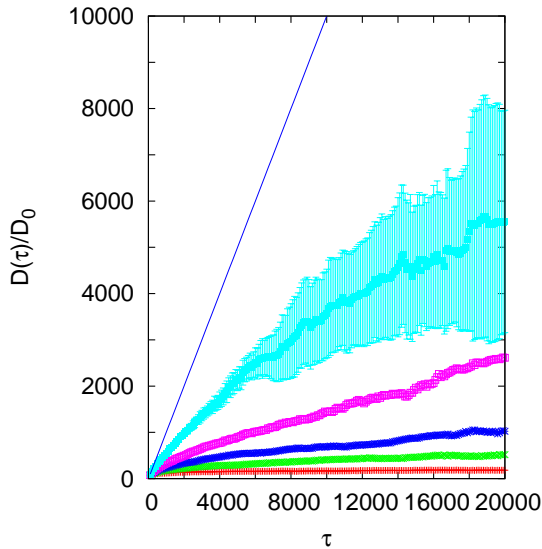} 
\caption{(Color online) 
Diffusion of the center of mass as a function of the Monte Carlo step index, $\tau$. All cases correspond to $s_2/E_{R_2}$ = 1.5 and different
values of $s_1/E_{R_1}$. From bottom to top: $s_1$ = 16,12,10,8,5. The solid line gives the slope corresponding to $\rho_s$/$\rho$ = 1.
The error bars are only displayed in one case for simplicity. }
\label{fig3}
\end{center}
\end{figure}

We calculated the superfluid fraction ($\rho_s$/$\rho$) using a variant of the winding number technique \cite{PhysRevLett.74.1500,PhysRevB.73.224515}. The formal 
expression employed was:
\begin{equation}
\frac{\rho_s}{\rho} = \lim_{\tau\to\infty}  \frac{D(\tau)}{ \tau D_0}
\label{super1}
\end{equation}
where $D_0=\hbar^2/2m$. $\tau$ indexes each Monte Carlo step of a given simulation, from the initial step after the Monte Carlo equilibration ($\tau$= 0)  
to the end of the Markov chain (in our case, $\tau$ = 20000). $D(\tau)$ is defined as:     
\begin{equation}
D(\tau) = \frac{1}{2 N } \langle \left(\vec{r}_{CM}(\tau)-\vec{r}_{CM}(0)\right)^2 \rangle \, 
\label{super}
\end{equation}
Here, $\vec{r}_{CM}(\tau)$ is the position of the center of mass of all the particles in the simulation cell at a step labeled by $\tau$. 
$N$ is, as before, the number of particles. This definition means that if we represent $D(\tau)/D_0$ as a function of $\tau$, 
for large enough $\tau$, we should get a straight line whose slope will give us the value of the superfluid fraction.  
An example of that procedure is displayed in Fig. \ref{fig3}. Each curve represents the average of 50 independent DMC simulations. 
There, $s_2/E_{R_2}$= 1.5 in all cases, while $s_1$ goes from $s_1/E_{R_1}$ = 16 (bottom) to $s_1/E_{R_1}$ = 5 (top). The solid line corresponds to the case
of a perfect superfluid ($D(\tau)/D_0$ = $\tau \rho_s/\rho =  \tau$), and it is included as a reference. The error bars are displayed only 
for the case $s_1/E_{R_1}$ = 5, are fairly typical, and correspond to an standard deviation of the data produced in those 50 independent simulations. What we can see is that 
the slope of $D(\tau)/D_0$ for large values of $\tau$ ($>$ 10000) increases from zero ($s_1/E_{R_1}$ = 16,12), to values that imply the existence of superfluids
($s_1/E_{R_1}$ = 10,8,5). 

\begin{figure}
\begin{center}
\includegraphics[width=0.44\textwidth]{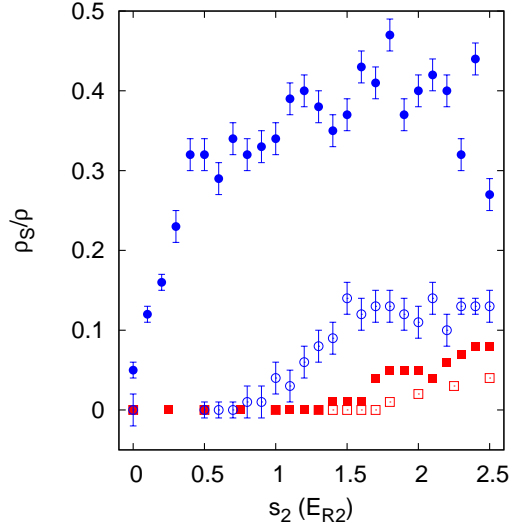} 
\caption{(Color online) Superfluid fraction as a function of $s_2$ (in units of $E_{R_2}$) for different $s_1$ values (in units of $E_{R_1}$). Open squares, $s_1/E_{R_1}$ = 16;
full squares, $s_1/E_{R_1}$ = 12; open circles, $s_1/E_{R_1}$ = 8; full circles, $s_1/E_{R_1}$= 5. The error bars are only displayed in the last two cases for simplicity, 
the ones not shown being of the same size ($\Delta(\rho_s/\rho) \sim$ 0.01).  
}
\label{fig4}
\end{center}
\end{figure}

From the slopes of curves similar to those in Fig. \ref{fig3}, and with the help of Eq. (\ref{super1}), we can deduce the superfluid 
fraction for the set of parameters we are interested in. Fig. \ref{fig4} shows those results for different
$s_1/E_{R_1}$'s as a function of $s_2/E_{R_2}$. The errors bars for each $\rho_s$/$\rho$ were calculated by the same method as in Ref. \onlinecite{feli3}: least-squares
fitting procedures were performed for the average values of $D(\tau)/D_0$ as a function of $\tau$, and for the same values after the subtraction (or
addition) of the corresponding limits given by the error bars of that ratio. This allows us to obtain a range of most probable slopes, and from that, a  
superfluid fraction interval. What we observe is that, except for the $s_1/E_{R_1}$ = 5 case, there is a range of $s_2/E_{R_2}$ values above which the system is an
insulator, and below  which we have a superfluid. We consider an arrangement to be non-superfluid when the
error bar for the superfluid fraction includes the value $\rho_s$/$\rho$ = 0. So, for instance, when $s_1/E_{R_1}$ = 12 (full squares in Fig. \ref{fig4}), the last
insulator corresponds to $s_2/E_{R_2}$= 1.6, since in that case $\rho_s$/$\rho$= 0.01 $\pm$ 0.01. 
This translates into the point $s_1/E_{R_1}$= 12, $s_2/E_{R_2}$ = 1.6 in Fig. \ref{fig2}. For larger $s_2/E_{R_2}$'s, we have a superfluid,  
and when  1.3 $< s_2/E_{R_2} < 1.6$, the phase is a 
gapless insulator. i.e. a Bose glass. The remaining of the full symbols in Fig. \ref{fig2} have been obtained in a similar way for other $s_1/E_{R_1}$ values. The error
bars displayed there have the same origin as the ones for the MI-BG limit. The full symbols in Fig. \ref{fig2} are related to the superfluid behavior of the system 
(see below).       

\subsection{Half a particle per potential well} 

\begin{figure}
\begin{center}
\includegraphics[width=0.44\textwidth]{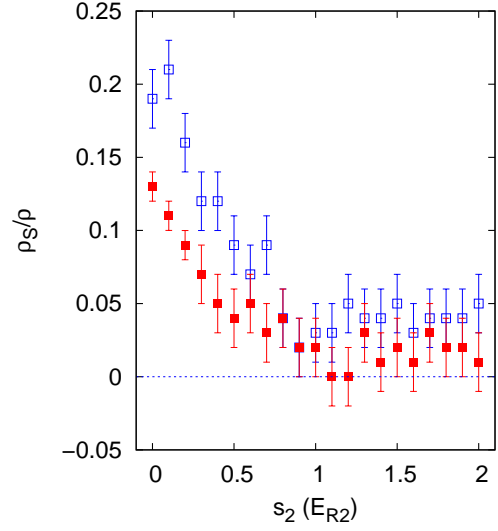} 
\caption{(Color online) Same as Fig. \ref{fig4}, but for a density of half a particle per potential well. Full squares, $s_1/E_{R_1}$ = 10; open squares, $s_1/E_{R_1}$ = 8.   
}
\label{fig5}
\end{center}
\end{figure}

\begin{figure}
\begin{center}
\includegraphics[width=0.44\textwidth]{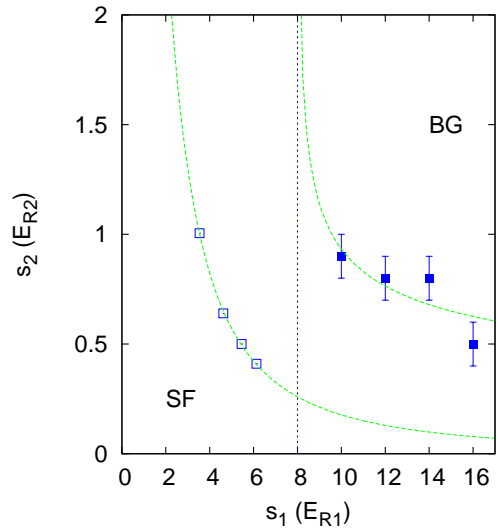} 
\caption{(Color online) Same as in Fig. \ref{fig2}, but for an average density of half a particle per potential well. The full symbols correspond to our simulation results
while the open symbols are from Ref. \onlinecite{roux}. All the lines displayed are guides-to-the-eye. See further explanation in the text.    
}
\label{fig6}
\end{center}
\end{figure}

Exactly the same calculations can be performed for an average density of half a particle per potential well. The only difference is that in the 
phase diagram we will have only superfluid and Bose glass phases, since a MI is only possible for a filling fraction of one. Fig. \ref{fig5} is 
the counterpart of Fig. \ref{fig4} for this lower density. There, we displayed the superfluid fraction as a function of $s_2/E_{R_2}$ for $s_1/E_{R_1}$ = 8 (upper curve)
and $s_1/E_{R_1}$ = 10 (lower curve), both sets of calculations including their corresponding error bars for each point. We can see that for 
$s_1/E_{R_1}$ = 8, the system is always a superfluid, irrespectively of the value of $s_2/E_{R_2}$. On the other hand, for $s_1/E_{R_1}$ = 10, an insulator  
(Bose glass) appears for $s_2/E_{R_2}\ge$ 0.9. This last value is represented in Fig. \ref{fig6} as the leftmost full symbol. Additional calculations
indicate that for this density no Bose glass appears when $s_1/E_{R_1} < 8$, at least for $s_2/E_{R_2}$'s in the range displayed in Fig. \ref{fig6}. 
For larger $s_1/E_{R_1}$'s, the limits between the superfluid and insulator phases are also shown as full symbols. The upper dashed line 
is a least-squares fit of those simulation results to a function of the form $s_2(s_1) = a/(s_1-8)^b$ ($a,b$, constants, $s_2$ and $s_1$ in units of their respective recoil energies), i.e., to a function with 
an  asymptote for $s_1/E_{R_1}$ = 8 (vertical dotted line), and it is intended simply as a guide-to-the-eye. 

On the other hand, the open symbols in Fig. \ref{fig6} 
correspond to the values obtained for a Bose-Hubbard Hamiltonian that models the same system \cite{roux}.
For instance, the rightmost open symbol of Fig. \ref{fig6} correspond to the largest $U/J$ value considered in Ref. \onlinecite{roux}, ($U/J$ = 10), 
for which the limit between the superfluid and the Bose glass phases is located at              
$s_2 E_{R_2}/J$ = 4. Using Eqs. \ref{J} and \ref{U} above, we have that this translates into $s_1/E_{R_1}$ = 6.1 and $s_2/E_{R_2}$ = 0.4.
A similar procedure was used to obtain the remaining open symbols shown there, the only cases whose transformations give coordinates in the $s_2/E_{R_2}$ range displayed in  Fig. \ref{fig6}. 
The dashed line on top of them is again a least-squares fit to be used only as a guide. What we observe is that both in our simulations and those of Ref. \onlinecite{roux}
predict a Bose glass for high enough $s_2/E_{R_2}$ and $s_1/E_{R_1}$, even though the Bose-Hubbard Hamiltonian underestimates the values of those parameters needed to create an insulator.      

\section{Discussion}

In this work, we described the behavior of a set of neutral atoms loaded in a quasi-one dimensional bichromatic optical lattice.    
To do so, we solved the Schr\"odinger equation corresponding to the continuous Hamiltonian in Eq. (\ref{hamiltonian}), with an external 
potential defined by Eq. \ref{fa}. The only approximation employed, besides the use of a stochastic method to obtain the ground state, 
was the description of the two-body interaction by a hard-spheres potential. Within this framework, we obtained the phase diagrams 
of Figs. \ref{fig2} and \ref{fig6}, as functions of the real parameters at two different atom densities. This makes
straightforward the comparison between theory and experiment. In particular, when $s_1/E_{R_1}$ = 16, our value for the MI-BG limit was $s_2/E_{R_2}$=1.3 $\pm$ 0.1, 
a value within the experimental range (0.5 $< s_2/E_{R_2}< $1.5) \cite{fallani}.    

However, the standard approximation to describe an optical lattice is the use of a Bose-Hubbard model. In a classic case of putting the cart before the horse,    
the fact that a set of atoms loaded in an monochromatic optical lattice is a very good experimental realization of a BH Hamiltonian 
for {\em some} values of the lattice defining parameters \cite{gre2,jaksch}, has been used as an excuse to describe {\em any} optical lattice  
in {\em any} circumstances by this model. The main advantage of this approach is that 
the BH model is relatively easy to use and its behavior is well understood; the main problem being that it is only an approximation
that breaks down even for monochromatic lattices with shallow potential wells, as it has been established both experimentally \cite{haller} and theoretically \cite{feli3,feli4}
for the description of the superfluid-Mott insulator transition.    

In addition, the study of a bichromatic lattice implies corrections for both the $J$ and $U$ parameters that would include the effect of the 
second standing wave and that there are usually not considered. This means that even in those cases in which the primary lattice is deep enough 
to be well described by a BH Hamiltonian, the description of the lattice is not exact. 
That compounded error would be the cause of the differences between the phase diagram for half a particle per potential well calculated with a continous 
and a BH Hamiltonian shown in Fig. \ref{fig6}.   
Unfortunately, to our knowledge, there are 
no experimental data to compare those diagrams to. We can only say that the  
stability zone of the superfluid phase is reduced with respect to the case of a continuous Hamiltonian, the same general trend found for monochromatic lattices 
\cite{feli3}.  

The situation is similar for a density of one particle per potential well.  
As indicated above, for $s_1/E_{R_1}$ = 16, 
there is a reasonable agreement between the position of the MI-BG transition predicted from our calculations and the experiment.
Regretably, we cannot compare any BH predictions to that value, since $s_1$ = 16 translates (using Eqs. (\ref{J}) and (\ref{U}), with the scattering length taken from Ref. \cite{pethick}) into
$U/J \sim$ 130,  a larger value than the ones considered in the literature \cite{roth,roux,deng}.       
On the other hand, 
the upper stability limit for the MI phase in Ref. \onlinecite{roux} for  $U/J =10$ (larger $U/J$ used) corresponds to $s_2 E_{R_2}/J \sim$ 10. 
This means $s_1/E_{R_1} \sim$ 6.1 and $s_2/E_{R_2} \sim$ 1.1, to
be compared to $s_1/E_{R_1}$ = 6, $s_2/E_{R_2}$ = 0.6 $\pm$ 0.2, from Fig. \ref{fig6}. Thus, we have that a BH description produces a shrinking of the SF stability zone with respect
to the case of a continuous model, as in the case of the lower density described above. This is corroborated by the onset of the SF-MI transition for $s_2$ = 0. 
The value in this work ($s_1/E_{R_1}$ = 5 $\pm$ 1) is larger than the one deduced in Ref. \onlinecite{roux} ($s_1/E_{R_1} \sim$ 3).

To conclude, we should make a last remark about the general form of the phase diagram displayed in Fig. \ref{fig2}. There, we can see that for some values of $s_1/E_{R_1}$ ($s_1/E_{R_1} >$ 10), an
increasing in $s_2/E_{R_2}$ makes the system change from a MI to a BG and further up, to a superfluid. However, when $s_1/E_{R_1}$ is smaller,  there is only a single transition, MI $\rightarrow$ SF.
This is in apparent contradiction to some previous theoretical results \cite{fisher,gurarie} for a Bose-Hubbard model, that indicate that between a MI and a SF phases, there should be always a BG "sandwich".
At least part of the discrepancy can be explained by the fact that those works correspond to systems with random site disorder, which behave differently than arrangements with 
quasiperiodic bichromatic lattices. In any case, the results of Fig. \ref{fig2} indicate that if that intermediate BG exists for $s_1/E_{R_1} <$ 10, the width of its stability interval should
be very narrow.

\acknowledgments
We acknowledge partial financial support from the 
Junta de Andaluc\'{\i}a group PAI-205 and grant FQM-5987, DGI (Spain) grant No. FIS2010-18356.


\end{document}